%% file: Fischer-Wimberger.tex
\documentclass[aps,prl,showpacs,twocolumn,amsmath,amssymb,superscriptaddress]{revtex4-1}
\usepackage{graphicx}
\usepackage{dcolumn}
\usepackage{bm}
\usepackage{units}
\usepackage{upgreek}
\usepackage{ucs}
\usepackage{color}
\usepackage{cancel}
\usepackage{hyperref}
\usepackage[all]{hypcap}
\usepackage{braket}

\input{input.tex}

\usepackage[english]{babel}

\begin{document}

\keywords{Bose-Hubbard model, quantum transport, ultracold atoms, open many-body quantum systems, many-body tunneling}
\title{Models for a multimode bosonic tunneling junction}
\author{David Fischer}
\affiliation{Institut f\"ur Theoretische Physik, Universit\"at Heidelberg, Philosophenweg 12, 69120 Heidelberg, Germany}
\author{Sandro Wimberger}
\affiliation{Institut f\"ur Theoretische Physik, Universit\"at Heidelberg, Philosophenweg 12, 69120 Heidelberg, Germany}
\affiliation{Dipartimento di Scienze Matematiche, Fisiche e Informatiche, Universit\`{a} di Parma, Parco Area delle Scienze 7/A, 43124 Parma, Italy}
\affiliation{INFN, Sezione di Milano Bicocca, Gruppo Collegato di Parma, Italy}

\begin{abstract}
We discuss the relaxation dynamics for a bosonic tunneling junction with two modes in the central potential well. We use a master equation description for ultracold bosons tunneling in the presence of noise and incoherent coupling processes into the two central modes. Whilst we cannot quantitatively reproduce the experimental data of the setup reported in [Phys. Rev. Lett. {\bf 115}, 050601 (2015)], we find a reasonable qualitative agreement of the refilling process of the initially depleted central site. Our results may pave the way for the control of bosonic tunneling junctions by the simultaneous presence of decoherence processes and atom-atom interaction.
\end{abstract}


\maketitle

\section{Introduction}
\label{sec:intro}

Since the first realization of Bose-Einstein condensates in 1995 \cite{BEC}, ultracold atoms have been providing vast opportunities for the investigating the quantum matter with an amazing experimental precision \cite{lewen07,bloch08}. Today, experiments can perform in situ measurements with lattice-site resolution \cite{singleS}. In particular, the use of an electron beam to remove ultracold atoms from a Bose-Einstein condensate and, more specifically, from selected sites in an optical lattice has opened up new possibilities to study non-equilibrium quantum transport\cite{nat_ott,prl09_ott,prl13_ott,prl15_ott,prl16_ott}. 

Inspired by the experimental results reported in ref.~\cite{prl15_ott}, we re-examine the effects of the dynamically changing number difference, interactions and dephasing on the filling of an initially empty central well in a three-site Bose-Hubbard model. The experiment was essentially two-dimensional with many accessible radial modes in a one-dimensional lattice. Moreover, a thermalization process was present which guaranteed a quasi steady-state flow into the initially depleted site. Such complications are hard to model within a many-body approach. In fact, the theory applied in ref. \cite{prl15_ott} is based on a discrete nonlinear Schr\"odinger equation with phase noise just in the initially depleted well. While very crude, this mean-field approach could describe the experimental data very well. However, the microscopical details of this thermalization process remain unclear in this approach. Therefore, it would be desirable to explain such results by a many-body theory as well, in particular in view of future experiments which may be done in the strongly correlated regime. Two attempts based on an inline three-site Bose-Hubbard model report seemingly contradicting results \cite{ann_phys_2015,comment}. Ref. \cite{comment} supports that the experimentally observed negative differential conductivity may solely be induced by a coherent tunneling mechanism. Yet, what seems crucial from the original analysis \cite{prl15_ott} as well as from the results reported in \cite{ann_phys_2015} is the presence of a dephasing process. Only such incoherent dephasing allows for the definition of a quasi-stationary current and suppresses coherent oscillations of the middle well population guaranteeing that the many-body dynamics saturates into an equilibrium steady-state, like it is found in the experiment.

\begin{figure}[tb]
\centering
\includegraphics[width=1.1\linewidth]{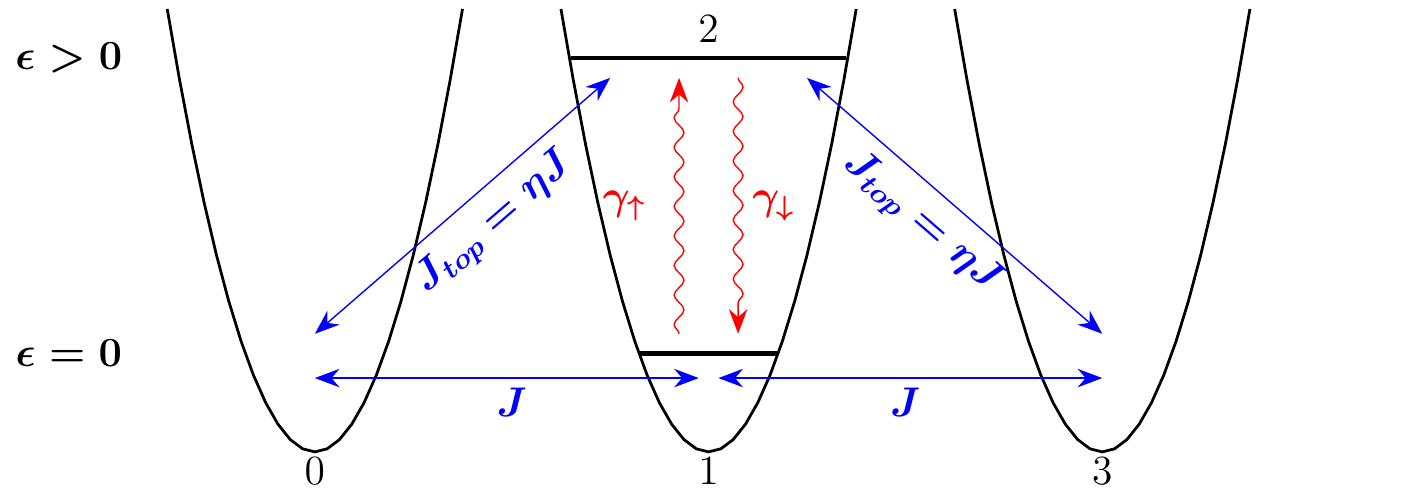}
\caption{Our model consisting of four modes, two of which represent the ground 1 and excited state 2 of the central well. The atoms tunnel from the modes 0 and 3 into the initially depleted middle well, coupling to both the ground and the excited mode, respectively. This coupling is effectively decreased by the difference in chemical potential for the ground state and by the reduction factor $\eta < 1$ for the excited state. While our models I and II assume just phase noise in the wells (not shown), model III assumes an incoherent coupling between the ground and excited state in the central well by corresponding relaxation and excitation processes sketched by the wiggly lines.}
\label{fig:1}
\end{figure}

Here, we present an extended model taking into account the presence of two modes in the initially depleted middle well. Such a scenario represents a minimal model of the radial modes in the actual experimental realization. Our basic setup is sketched in Fig. \ref{fig:1} where the modes 1 and 2 represent the two modes of the central well. The other two wells are coupled to both of these modes but contain themselves just one mode each. The coherent evolution is defined by the following many-body Bose-Hubbard Hamiltonian
\begin{eqnarray}
& H  =  \nonumber \\ 
&\sum_{j=0}^3 \epsilon^{}_j n^{}_j + \frac{U}{2}\sum_{j=0}^3 n^{}_j(n^{}_j-1) - \frac{1}{2} \sum_{j=0,3} J ( a^\dag_j a^{}_{1} +  a^\dag_1 a^{}_{j}) \nonumber \\ 
& - \frac{1}{2} \sum_{j=0,3} J_{\rm top} (a^\dag_j a^{}_{2} + a^\dag_2 a^{}_{j})
\label{eq:bose-hubbard}
\end{eqnarray}
where $n_j = a^\dag_j a^{}_j$ is the number operator, and $a^\dag_j, a^{}_j$ are the creation and annihilation operators for the $j$-th site of the periodic lattice, obeying the bosonic commutation relation $[a^{}_j,a_i^\dagger]=\delta_{ij}$. We set $\hbar = 1$, measuring all energies in frequency units. In what follows, we will use $\epsilon_{0,1,3}=0$ and only $\epsilon_{2}$ will be positive in order to model a higher lying mode in the middle well. We also set $J=1$, such that energies are expressed in units of $J$ and times in units of $J^{-1}$. The hopping matrix element to the excited state is reduced to $J_{\rm top}=\eta J=\eta$ in order to model a smaller coupling to the upper mode. This corresponds to the experimental situation reported in ref. \cite{prl15_ott}, where $0< \eta <1$ would be the Frank-Condon factor which slightly suppressed the coupling to excited modes.

We assume the relaxation time of the bath -- which may be a thermal gas cloud surrounding the condensate -- to be much smaller than the typical timescale of our system $\tau_S$. Additionally to this so called Markov-condition, we use the Born approximation and consider only weak interactions with the bath, such that the density matrix of the bath and the system separate. The physical condition to be verified for the latter is: 
\begin{equation}
	\tau_S \sim (NJ)^{-1} = \frac{1}{20} J^{-1} \ll \tau_R, 
\end{equation}
where the relaxation times $\tau_R$ are of the order of $J^{-1}$, please see below, that justifies the approximation.

The full quantum evolution is then described by the following master equation in Lindblad form \cite{daley14,epj_st_2015}
\begin{equation}
\dot{\varrho} = - \frac{i}{\hbar} [H,\varrho] + \mathcal{L}[\varrho]
\label{eq:master}
\end{equation}
with incoherent parts
\begin{equation}
\mathcal{L}[\varrho] = \sum^{}_{j} \gamma^{}_j \left( A^{}_j \varrho A_j^\dag - \frac{1}{2} A_j^\dag A^{}_j \varrho - \frac{1}{2} \varrho A_j^\dag A^{}_j \right) \,.
\label{eq:lindblad}
\end{equation}
The $A_j$ are the Lindblad operators and represent noise and relaxation processes specified below. The $\gamma_j$ give the rates, at which those processes occur.

\section{Numerical unraveling of the master equation by quantum jump method}
\label{sec:2}

Unfortunately, for more than two modes neither the coherent dynamics induced by the Hamiltonian (\ref{eq:bose-hubbard}) nor the master equation (\ref{eq:master}) can be solved analytically in general. Our method of choice to evolve the master equation in time is by calculating many individual quantum trajectories independently and then approximate the desired expectation values of observables $\hat{O}$ by averaging over the realizations $\ket{\psi^{(i)}(t)}$ \cite{bre06,daley14,epj_st_2015}.
\begin{equation}
\expval{\hat{O}}(t) = \Tr(\hat{O}\varrho(t)) = \lim\limits_{R \rightarrow \infty}
{\frac{1}{R} \sum_{i=1}^R \expval{\hat{O}}{{\psi}^{(i)}(t)} }
\label{eq:expectation}
\end{equation}
We can estimate the statistical error resulting from the finite number of realizations $R$ by calculating the standard deviation from the average. Taking this into account, we are able to choose $R$ accordingly to the desired precision
\begin{equation}
\sigma^2(R,t) = \frac{1}{R(R-1)}\sum_{i=1}^R \big(\hat{O}^{(i)}(t)-\overline{O}(t)\big)^2 \sim \frac{1}{R}\,.
\label{eq:error}
\end{equation}
In all our applications the propagation of $\ket{\psi^{(i)}}$ will be a peace-wise deterministic process, where the development of the state vector is caused by an effective non-hermitian Hamiltonian
\begin{equation}
H^{}_{\text{eff}}= H - \frac{i}{2} \sum_{j} \gamma^{}_j A^{}_jA_j^\dag \,,
\label{eq:qjump_ham}
\end{equation}
interrupted by sudden quantum jumps -- induced by the projection with the related Lindblad operator 
$$\ket{\psi'}=\frac{A_j\ket{\psi}}{\|A_j\ket{\psi}\|}$$  
whenever the decreasing norm of $\ket{\psi^{(i)}}$ reaches a given random value $\chi$. The threshold values $\chi$ are drawn from a uniform distribution in $[0,1]$. Each time the norm of the wave function reaches the threshold the new value is randomly drawn. This artificial reduction of the norm only serves to determine the exact moment of the next jump and will be removed afterwards by renormalizing the state.

In all applications below we keep the system of reasonable size in order to reduce the computational load over the entire number of trajectories. Hence we restrict to $N=20$ bosons which are initially equally distributed in the wells 0 and 3. As already mentioned $J=1$ is fixed as well as the other energy scales $\eta=0.3$ and $U=2$. The offset of the excited mode in the central well 2 is chosen such that it is resonant with the loss of one atom from the initially filled sites, i.e. $\epsilon=\frac{U}{2}(n(n-1)-(n-1)(n-2))=9U$ for $n=N/2=10$. 

We concentrate on showing the fraction of atoms in the lower mode of the central well, indexed by 1, as compared with the fillings in the sites 0 or 3 which are equal on average due to the symmetry of our setup. Since all populations in the modes are time-dependent we look at the time-dependent ratio $f(t) \equiv \frac{n_1(t)}{n_0(t)}$, which we call the normalized filling in ground mode 1 of the central site. With this definition a value of $f(t)=1$ indicates equal filling in the three ground modes of the model.

\section{Three dynamical models}
\label{sec:3}

In the following we will study three dynamical models which differ in the incoherent processes of the master equation. The initial state we pick for our time-evolution is in the self-trapped regime of the Bose-Hubbard model, meaning that without any kind of incoherent process no dynamics would occur on experimentally reasonable timescales. While the first two models I and II assume pure phase noise either in all or just in the middle well, model III assumes a biased coupling between the upper and lower states in the middle well. Whilst, for proper choices of the Liouville parameters, all of our models effectively lead to a steady-state saturation of the filling $f(t)$, they differ in the shape and the time scales of the relaxation to the new equilibrium state. We now start with the discussion of model I.

\subsection{Global phase noise}
\label{sec:3.1}

In a first approach we assume that the decoherence is induced by a global noise process, acting on each site with the same rate $\kappa$. The corresponding Lindblad operators are the particle number operators of the sites $A_j=n_j=A_j^\dag$, giving
\begin{equation}
\mathcal{L}[\varrho]=\frac{\kappa}{2}\sum_{i=0}^3 (2n_i\varrho n_i - n_i n_i\varrho - \varrho n_i n_i) \,.
\label{eq:liov1}
\end{equation}

We may optimize the maximally allowed time step $\delta t$ at which the norm of our trajectories $\|\psi^{(i)}\|$ is evaluated. It has to be much smaller than the typical time between two quantum jumps \cite{bre06}, meaning that
\begin{equation}
\frac{1}{k\expval{\sum_{i=0}^3 n_j^2}} \overset{}{\underset{n_j\leq N/2 = 10}{\geq}} \frac{1}{200\kappa} \gg \delta t \,,
\label{eq:tstep}
\end{equation}
in good approximation.

\begin{figure}[t]
\centering
\includegraphics[width=\linewidth]{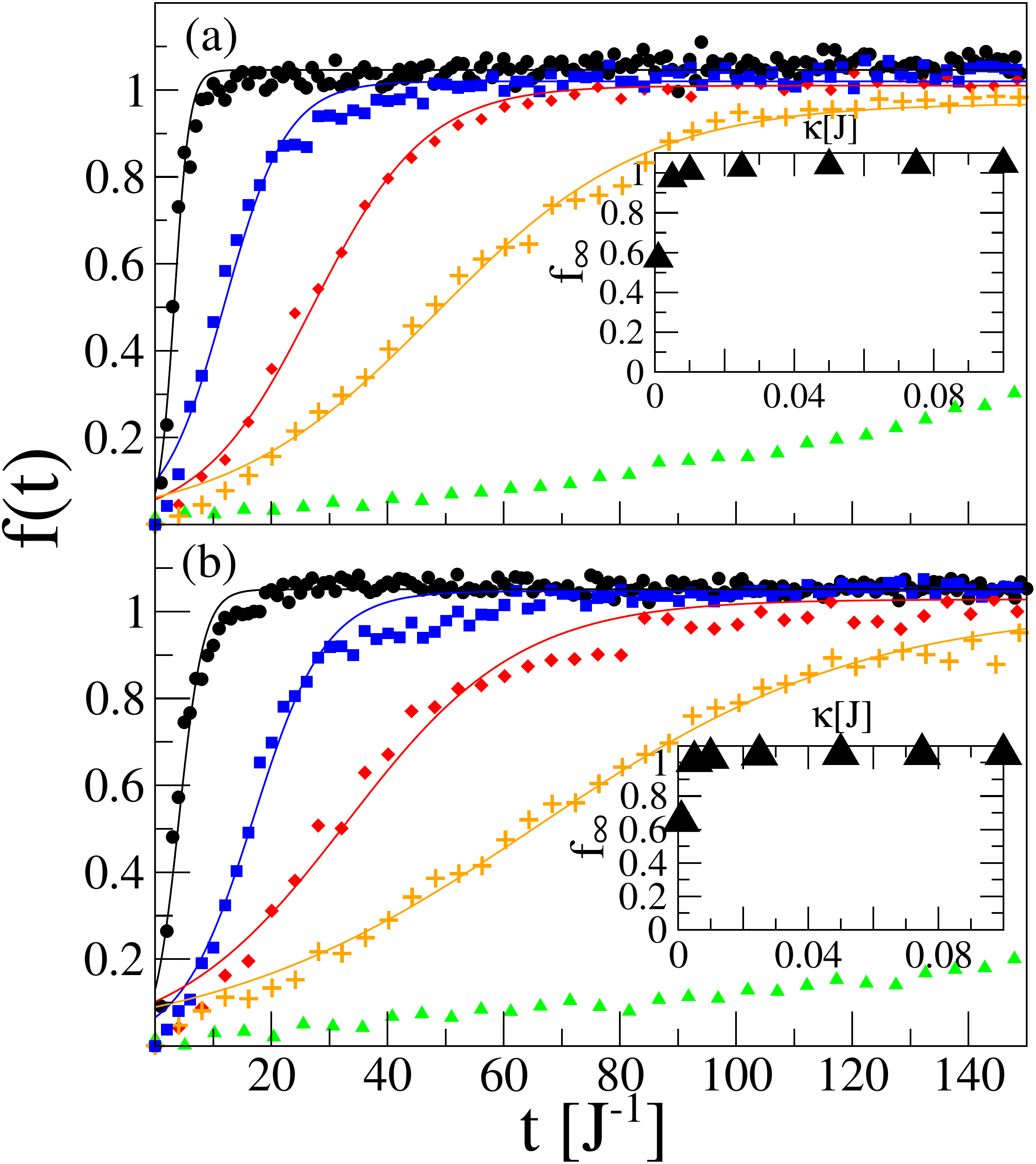}
\caption{Global noise, model I: normalized filling of ground mode of the central well $f(t)$ for different noise strength: $\kappa=0.1$ (black circles), $0.025$ (blue squares) $0.01$ (red diamonds), $0.005$ (orange plusses), 0.001 (green triangles). All curves result from an average over $\sim 100$ trajectories. The solid lines correspond to the fits according to Eq. (\ref{eq:n_fit}). The symbols in the insets display the fit parameter $f_\infty$ as a function of the noise amplitude 
$\kappa$. (a) for the setup sketched in Fig. \ref{fig:1} with two modes in the central well; (b) for just the ground mode in the central well.
}
\label{fig:2}
\end{figure}

Figure \ref{fig:2} collects our results for the normalized filling $f(t)$ of the ground state of the middle well for various values of $\kappa$. These curves show a characteristic s-form, already observed in the original experiment \cite{prl15_ott}, and are hence best fitted with a sigmoid function of the similar form
\begin{equation}
	g(f_\infty,\lambda,\tau;t) \equiv \frac{f_\infty}{1+e^{\lambda(t-\tau)}} \,,
	\label{eq:n_fit}
\end{equation}
where $f_\infty$ is the fit parameter describing the relative filling for $t \rightarrow \infty$ and $\tau$ gives the time when half of that filling is reached. $\lambda > 0$ is another fit parameter describing the maximal rate of change in $f(t)$. The insets Fig. \ref{fig:2}  show the dependence of $f_\infty$ on $\kappa$. Weak noise with $\kappa \leq 0.005$ already leads to a saturation of $f_\infty \approx 1$. For comparison we actually computed the evolution for two different systems, in Fig. \ref{fig:2}(a) for all four modes (0,1,2,3) and in Fig. \ref{fig:2}(b) consisting of just the three modes (0,1,3). 

\begin{figure}[t]
	\centering 
	\includegraphics[width=\linewidth]{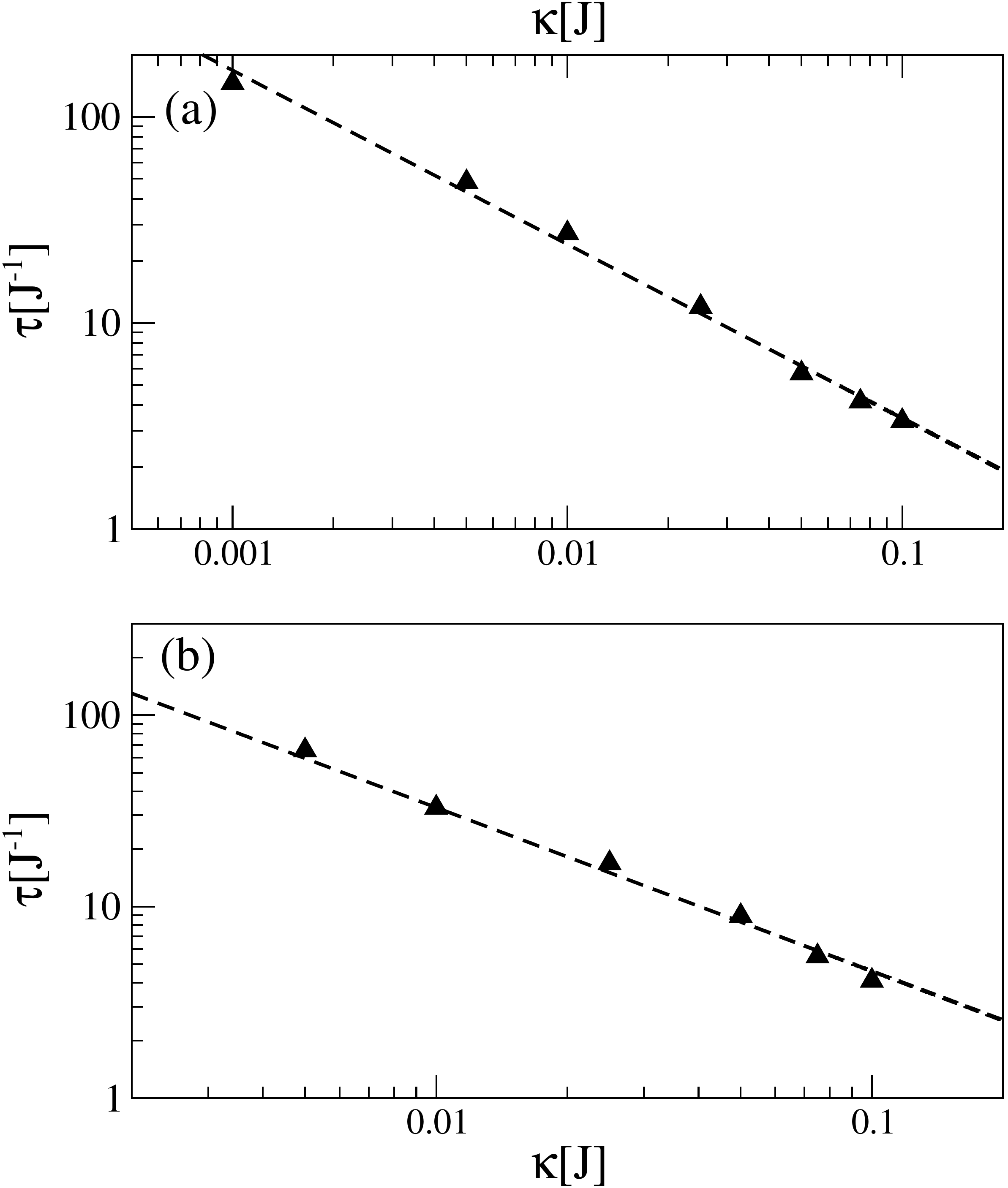}
	\caption{The symbols show the scaling of the refilling time with increasing noise strength $\kappa$ from the data shown in Fig. \ref{fig:2}, according to the fit procedure given in the main text. (a) from Fig. \ref{fig:2}(a) with two modes in the central well; (b) from Fig. \ref{fig:2}(b) for just one mode in the central well.}
	\label{fig:3}
\end{figure}

Interestingly, we find a power-law scaling of the refilling times $\tau$ with respect to the noise amplitude $\kappa$
\begin{equation}
	\tau(\kappa) \propto \kappa^{-\alpha} \,.
	\label{eq:power_fit}
\end{equation}
The exponents are $\alpha \approx 0.84$ for (a) and $\alpha \approx 0.85$ for (b), as seen in Fig. \ref{fig:3}. 

Overall, we can say that the addition of the second excited state mode in the middle well has qualitatively little effect on the evolution of the relative fillings $f(t)$. In other words, our new model with four modes gives similar results as the three-site Bose-Hubbard model analyzed in ref. \cite{ann_phys_2015} corresponding to what is seen in Fig. \ref{fig:2}(b).

\subsection{Local noise for the excited mode in the middle well}
\label{sec:3.2}

Our model II is similar to the previous one, but because the excited mode spreads wider in real space we can assume that it offers a bigger target for collisions with rest gas atoms or is more sensitive to fluctuations of the potential, for instance. By assuming that these collisions are the dominant factor, we neglect noise terms in all other modes. The resulting Liouvillian is then just
\begin{equation}
\mathcal{L}[\varrho]=\frac{\kappa}{2} (2n_2\varrho n_2 - n_2 n_2\varrho - \varrho n_2 n_2)\,.
\label{eq:liov2}
\end{equation}

\begin{figure}[tb]
	\centering
	\includegraphics[width=1\linewidth]{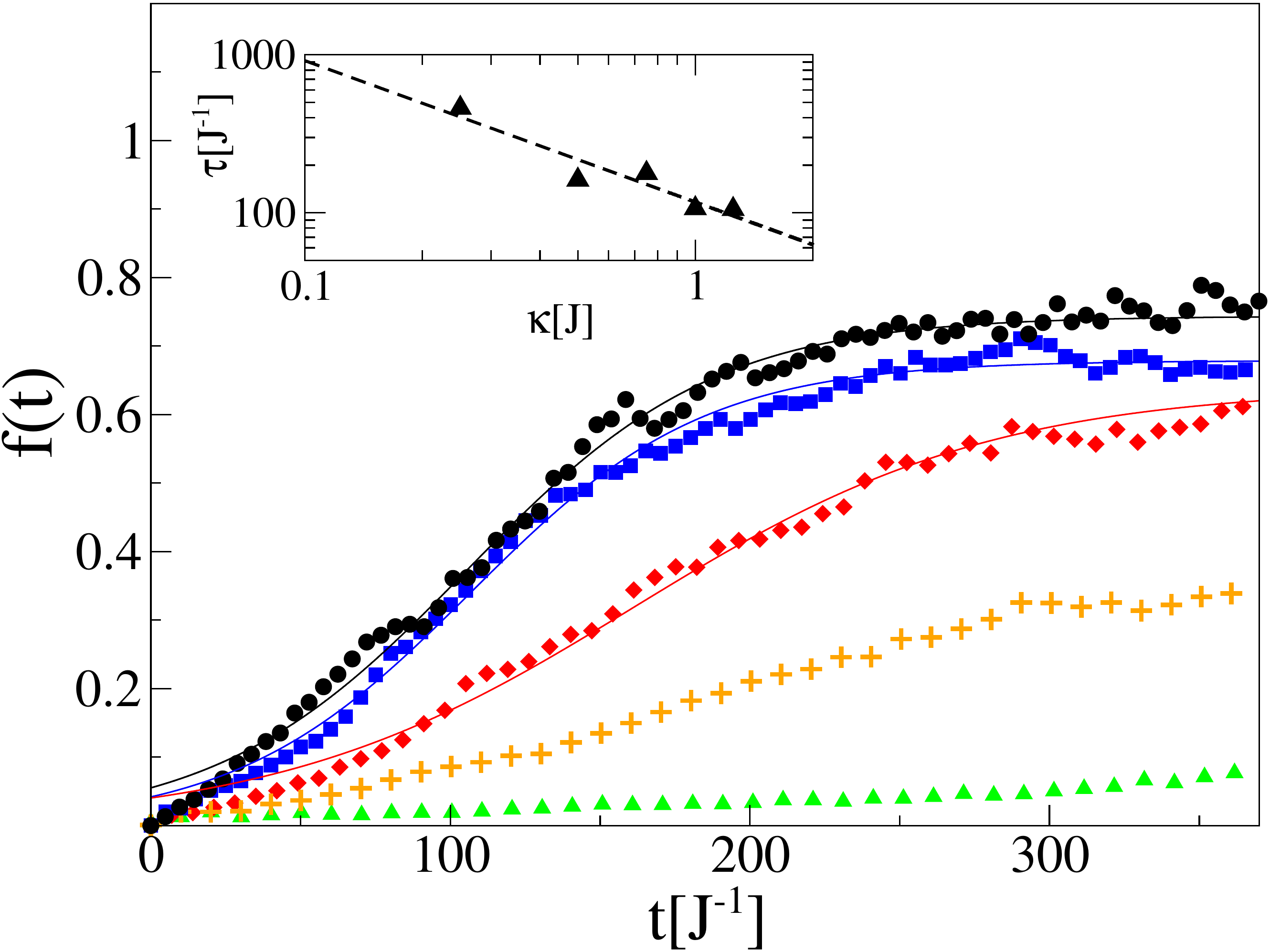}
	\caption{Local phase noise, model II: normalized filling of ground mode of the central well $f(t)$ for $\kappa=0.1$ (green triangles), $0.25$ (orange plusses) $0.5$ (red diamonds), $1$ (blue squares), 1.25 (black circles). The solid lines correspond to fits according to Eq. (\ref{eq:n_fit}) for the cases of stronger noise where the signal has started to saturate. The insets show the refilling times $\tau$ as a function of the noise strength $\kappa$ (black triangles), together with a power-law fit given by the dashed line.}
	\label{fig:4}
\end{figure}

Figure \ref{fig:4} presents the results obtained for the parameters stated above in section \ref{sec:2}. In this case the normalized population $f(t)$ never goes to one but at most saturates at values below 0.8 for reasonable time scales, even for relatively strong phase noise with $\kappa > 1$. In order to observe a substantial refilling the noise must be stronger than for our model I above. Since the saturation is hardly ever complete and reaches values less than one, the fits using (\ref{eq:power_fit}) are less good. Nevertheless the refilling time seems to scale also in this case in a power-law fashion with exponent $\alpha \approx 0.89$, see Fig. \ref{fig:4}. 

Finally, we note that applying the noise at the ground mode 1 does not help the refilling at all. The reason is that this mode is initially empty and the effect of the noise is small. Because of the great difference in chemical potential with respect to the initially full modes 0 and 3 the ground mode remains also close to zero filling if the noise {\em only} acts on mode 1.  To summarize, what we learn from our models I and II is that only the noise in the modes 0 and 3 or in the mode 2, respectively, really helps to enhance the refilling of the ground mode 1. 

\subsection{Relaxation in the middle well}
\label{sec:3.3}

Our final model III consists of explicit couplings between the two modes of the central potential well. Since the final state should relax in the ground state, we choose a biased incoherent coupling with two rates $\gamma_\downarrow, \gamma_\uparrow > 0$. As we will see, a necessary condition for a final state of $f_\infty \approx 1$ is $\gup < \gdo$. The Liouvillian then reads with $A^{}_\uparrow=a^{}_1 a_2^\dagger$ and $A_\downarrow =a^{}_2 a_1^\dagger$ with $A_\uparrow^\dagger=A^{}_\downarrow$:
\begin{align}
\mathcal{L}[\varrho]=\frac{\gamma_\downarrow}{2} (2 A_\downarrow \varrho A_\uparrow - A_\uparrow  A_\downarrow\varrho - \varrho A_\uparrow A_\downarrow) \nonumber\\
+\frac{\gamma_\uparrow}{2} (2 A_\uparrow \varrho A_\downarrow - A_\downarrow  A_\uparrow\varrho - \varrho A_\downarrow A_\uparrow)
\label{eq:liov3}
\end{align}

As in subsection \ref{sec:3.1} we can estimate that the integration time step $\delta t$ must be much smaller then the time scales given by the coupling processes, hence
\begin{align}
&\frac{1}{\gamma_\uparrow \expval{A_\uparrow A_\downarrow}+\gamma_\downarrow \expval{A_\downarrow A_\uparrow}} =\nonumber \\ 
&\frac{1}{\gamma_\uparrow(n_{2}+1)n_{1}+\gamma_\downarrow(n_{1}+1)n_{2}} \gtrsim \frac{1}{(\gamma_\uparrow+\gamma_\downarrow) 2 \cdot 10} \gg \delta t\,.
\label{eq:tstep3}
\end{align}
The last step assumes that the excited state 2 is never populated by more than two atoms, what we learn from the analysis of the numerical evolutions.

\begin{figure}[tb]
	\centering
	\includegraphics[width=\linewidth]{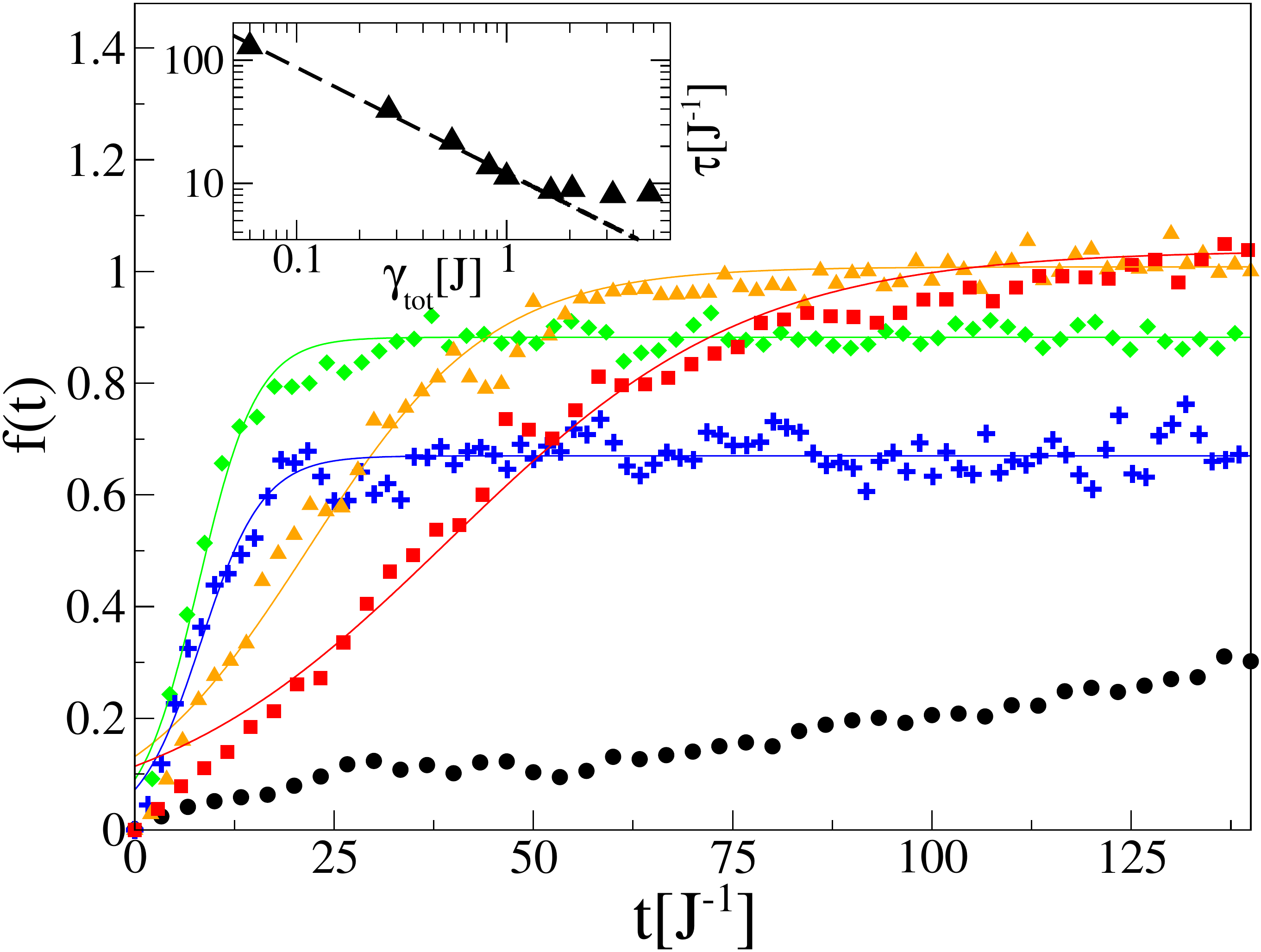}
	\caption{Incoherent mode couplings, model III: 
	normalized filling $f(t)$ for the following pairs of coupling rates: $(\gup,\gdo)=(0.01,0.05)$ (black circles),  
	$(0.025,0.25)$ (red squares), $(0.05,0.5)$ (orange triangles), $(0.02,3)$ (green diamonds), and $(0.3,4.5)$ (blue plusses).
	The solid lines correspond to fits according to Eq. (\ref{eq:n_fit}).
	The inset shows the refilling times vs the total rate $\gamma_{\rm tot}=\gdo + \gup$ (triangles) with power-law fit (dashed line, fitted up to
	 $\gamma_{\rm tot}=1$).}
	\label{fig:5}
\end{figure}

\begin{figure}[tb]
	\centering
	\includegraphics[width=\linewidth]{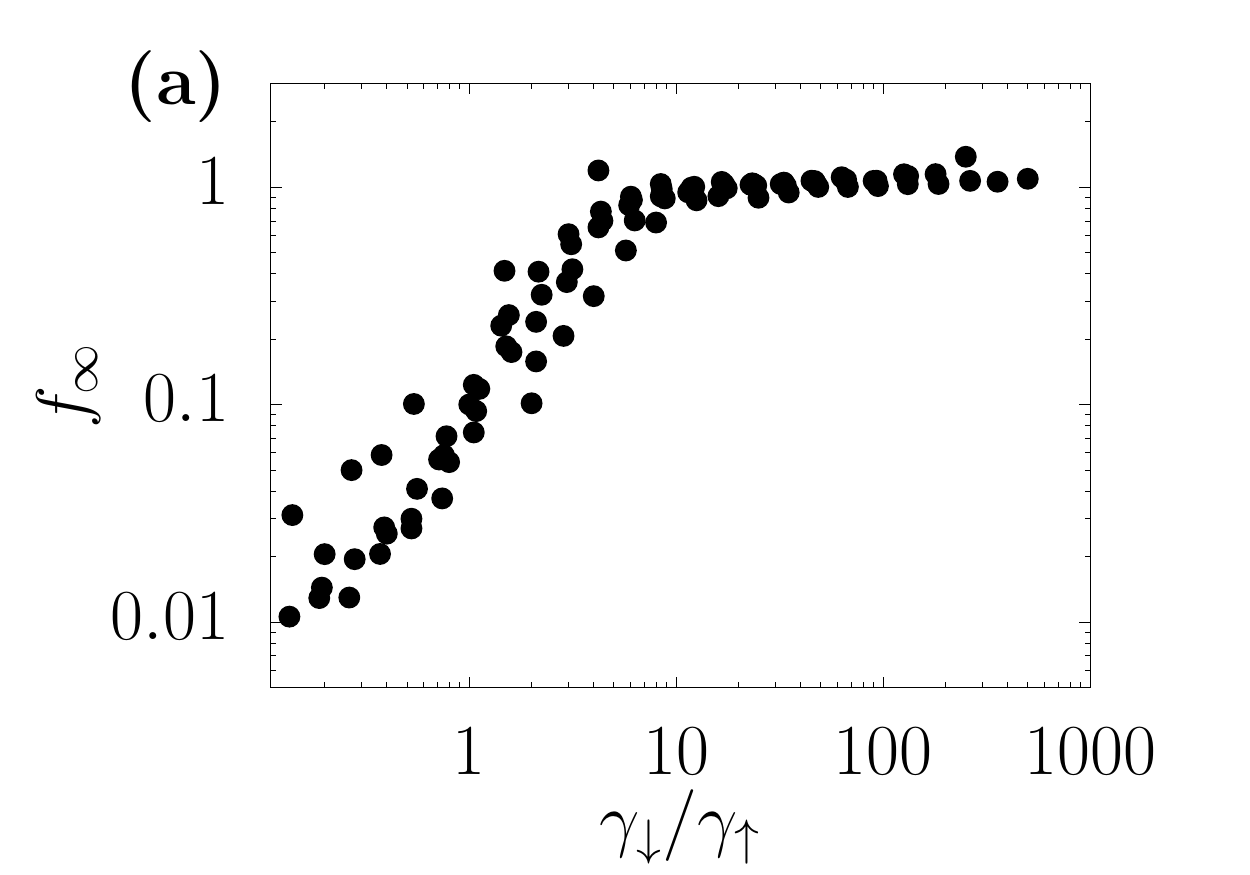}
	\vspace{2mm}
	\includegraphics[width=\linewidth]{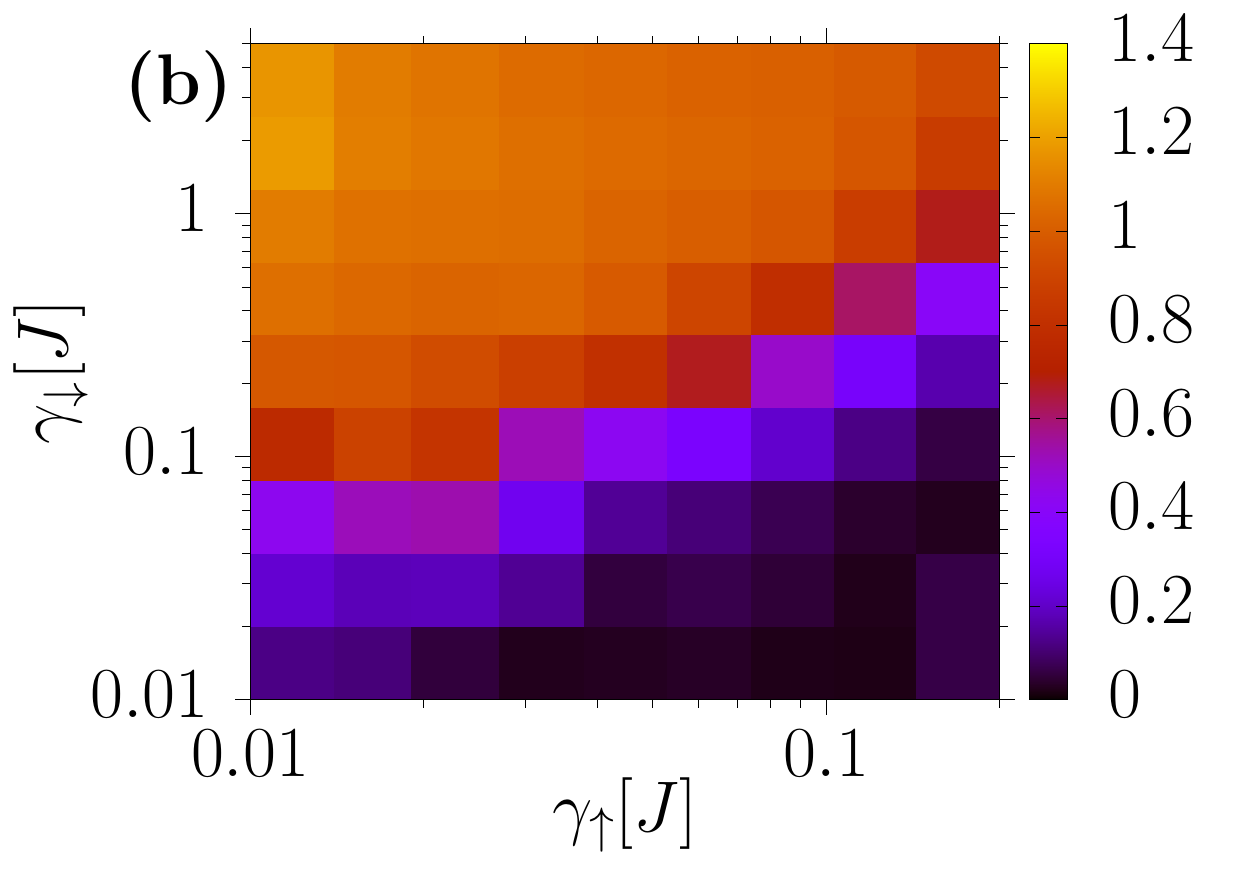}
	\caption{(a) Final filling $f_\infty$ as function of the ratio $\gdo / \gup$. We see a saturation for $\gdo / \gup \gtrsim 10$ at values close to one.
	(b) False color plot of the final filling $f_\infty$ as function of both $\gup$ and $\gdo$. As expected, the ground level 1 fills substantially if
	$\gdo$ dominates, while it remains essentially empty for larger $\gup \gtrsim \gdo$.
	}
	\label{fig:6}
\end{figure}

The results of the filling process are reported in Fig. \ref{fig:5} for various pairs of parameters $(\gup,\gdo)$. On average the normalized filling $f(t)$ saturates. The saturation value is around one for ratios of $\gdo / \gup \gtrsim 10$, see Fig. \ref{fig:6}(a). A further analysis of the asymptotic filling $f_\infty$ over the plane of parameters $(\gup,\gdo)$ is given in Fig. \ref{fig:6}(b). 

Extracting the refilling times based on Eq. (\ref{eq:power_fit}) we find that these times scale with the total rate $\gamma_{\rm tot}=\gdo + \gup$ as shown in the inset of Fig. \ref{fig:5} . The scaling is again power-law like with exponent $\alpha \approx 0.86$ until the total rate becomes comparable with the hopping matrix element $\gamma_{\rm tot} \approx J=1$. For $\gamma_{\rm tot} > J$, the refilling times remain approximately constant not depending any more on the strength of the incoherent coupling processes, which then dominate anyhow in this situation. Additional weak phase noise, either globally (model I) or locally (model II) acting, on average changes little in the obtained results as we tested for exemplary cases (not shown here). 

\section{Conclusions}
\label{sec:con}

Inspired by the experiment reported in ref. \cite{prl15_ott}, we have studied minimal models for a similar system of strongly interacting bosons. All our models I-III are based on an incoherent mechanism in order to overcome self-trapping as well as to avoid oscillatory behavior of the populations between modes, which both are not observed experimentally. Both, appropriate phase noise as well as biased couplings between the two states of the initially depleted central well, predict an increase of the middle well population with a saturation characterizing the equilibrium situation. The latter equilibrium is reached, however, at times, we called them refilling times, which in detail depend on the precise parameters of our models. We confirm the importance of incoherent processes in the dynamics, in accordance with the much simpler model used in \cite{prl15_ott}. Remarkably, in all the cases studied here the refilling times scale algebraically over a wide parameter range with the rates of the corresponding incoherent process, and even with comparable power-law exponents $\alpha \sim 0.8 \ldots 0.9$. This prediction might be verified  in an experiment with controllable noise, as an example of a thermalization process which shows -- similar to a phase transition -- a diverging timescale at vanishing noise.

Possible further extensions of such models could include more modes in the central site as well as in the initially filled sites. Such models are topologically similar to two-dimensional Bose-Hubbard problems which have been investigated only recently, see e.g. \cite{2DBHM} for closed models and \cite{2Dopen} for systems with dissipation present. All such setups are primers for the study of quantum transport in a many-body as well as an open system's context \cite{prl15_ott,QTexp}, with possible applications to atomtronics \cite{Qtrans}.

\section*{Acknowledgments}
First we thank Herwig Ott and his group for sharing many insights on the experimental implementation with us.
Moreover, we are very grateful to Elmar Bittner for computational and logistic support at the ITP. 
D. F. acknowledges support by the Studienstiftung des deutschen Volkes.

\end{document}

%% file: input.tex
\usepackage{todonotes}
\usepackage{amsmath}
\usepackage{color}

\usepackage{graphicx}
\usepackage{latexsym}
\usepackage{amsfonts}
\usepackage{hyperref}
\hypersetup{colorlinks=false, pdfborder={0 0 0}}
\usepackage[english]{babel}
\usepackage{bm}
\usepackage{units}
\usepackage{upgreek}
\usepackage{color}
\usepackage{tabularx}
\usepackage{soul}
\usepackage{lipsum}
\usepackage{braket}
\usepackage{physics}
\usepackage{cleveref}

\newcommand{\gup}{\gamma_{\uparrow}}
\newcommand{\gdo}{\gamma_{\downarrow}}

\mathchardef\ordinarycolon\mathcode`\:
\mathchardef\ordinaryequal\mathcode`\=
\mathcode`\:=\string"8000
\begingroup \catcode`\:=\active
  \gdef:{\mathrel{\mathop\ordinarycolon}}
\endgroup
\begingroup \catcode`\==\active
  \gdef={\mathrel{\mathop\ordinaryequal}}
\endgroup